\documentclass[%
 twocolumn,
 superscriptaddress,
 preprintnumbers,
 nofootinbib,
 amsmath,amssymb,
 aps,
 prl,
]{revtex4}
\usepackage{psfrag,slashed,cancel,array,graphicx,hyperref}
\usepackage{subfigure}
\usepackage[utf8]{inputenc}
\usepackage{mathtools}
\usepackage{mathrsfs}
\usepackage{multirow}
\usepackage{braket}
\usepackage{booktabs} %nice tables
\usepackage[normalem]{ulem}
\hypersetup{pdftitle={},pdfcreator={},linkcolor=[rgb]{0.15,0.35,0.75},colorlinks=true,citecolor=[rgb]{0.675,0,0.2},urlcolor=[rgb]{0.15,0.35,0.65}}

\newcommand{\nn}{\nonumber}

\def\xt{\tilde{x}}
\def\xb{\bar{x}}
\def\logx{l_x}
\def\logxbar{l_{\bar{x}}}
\def\lithxbar{\,\text{Li}_3(\bar{x})}
\def\litwxbar{\,\text{Li}_2(\bar{x})}
\def\sotxbar{\,\text{S}_{12} (\bar{x})}

\def\iomx{\frac{1}{\bar{x}}}

\def\zt{\tilde{z}}
\def\zb{\bar{z}}
\def\logz{l_z}
\def\logzbar{l_{\bar{z}}}
\def\lithzbar{\,\text{Li}_3(\bar{z})}
\def\litwzbar{\,\text{Li}_2(\bar{z})}
\def\sotzbar{\,\text{S}_{12} (\bar{z})}

\def\iomz{\frac{1}{\bar{z}}}

\allowdisplaybreaks

\begin{document}

%\preprint{DESY}

\title{NNLO QCD corrections to semi-inclusive DIS}

\author{Saurav Goyal}
\email{sauravg@imsc.res.in}
\affiliation{The Institute of Mathematical Sciences,  A CI of Homi Bhabha National Institute, Taramani, 600113 Chennai, India}
\author{Sven-Olaf Moch}
\email{sven-olaf.moch@desy.de}
\affiliation{II. Institute for Theoretical Physics, Hamburg University, D-22761 Hamburg, Germany} %Luruper Chaussee 149,
\author{Vaibhav Pathak}
\email{vaibhavp@imsc.res.in}
\affiliation{The Institute of Mathematical Sciences,  A CI of Homi Bhabha National Institute, Taramani, 600113 Chennai, India}
\author{Narayan Rana}
\email{narayan.rana@niser.ac.in}
\affiliation{School of Physical Sciences, National Institute of Science Education and Research, An OCC of Homi Bhabha National Institute, 752050 Jatni, India}
\author{V. Ravindran}
\email{ravindra@imsc.res.in}
\affiliation{The Institute of Mathematical Sciences,  A CI of Homi Bhabha National Institute, Taramani, 600113 Chennai, India}

\date{\today}

%----------------------------------------------------------------------------------------------------%             
\begin{abstract}
 We present the first results for the next-to-next-to leading order (NNLO) corrections to the semi-inclusive deep-inelastic scattering process in perturbative quantum chromodynamics. 
 We consider the quark initiated flavor non-singlet process and obtain the complete contributions analytically at leading color.
 All relevant virtual and real emission Feynman diagrams have been computed  
 using integration-by-parts reduction to master integrals
 and two approaches for their subsequent evaluation (parametric phase-space integration and method of differential equations).
 The numerical analysis demonstrates the significance of the NNLO corrections and their great impact on the reduction of the residual scale dependence.
\end{abstract}
%----------------------------------------------------------------------------------------------------%

% PACS, the Physics and Astronomy Classification Scheme.
% \keywords{SIDIS}

\maketitle

%----------------------------------------------------------------------------------------------------%
% \section{Introduction}
Landmark inclusive measurements of structure functions (SF) in deep-inelastic scattering (DIS) experiments provide
a wealth of information on the internal structure of hadrons at high energies in terms of quarks and gluons and give valuable insights into the underlying strong interaction dynamics. 
Collinear factorization in quantum chromodynamics (QCD) with the systematic separation of short- and long-distance phenomena provides the theoretical foundations for the use of SF data in the extraction of the process-independent parton distribution functions (PDFs), which describe the parton dynamics within the colliding hadrons.
Semi-inclusive DIS (SIDIS) measurements with an observed hadron in the final state allows, in addition, for the study of the parton (quark/gluon) to hadron fragmentation, encoded in universal fragmentation functions (FFs). 
This makes SIDIS the most promising and valuable probe of PDFs and FFs at the upcoming Electron-Ion collider (EIC) at the Brookhaven National Laboratory in the USA.
The EIC opens up unique opportunities to explore the nucleon structure as well as the underlying dynamics of hadrons in various environments.
This includes a thorough understanding of the light-quark flavor PDFs and also the spin structure of the nucleon, using polarized beams, to measure the polarized PDFs.
A quantitative assessment of the anticipated precision of SIDIS measurements and their impact on the determination of various hadronic observables at the EIC has been presented in ref.~\cite{Aschenauer:2019kzf}.

We consider the SIDIS process  
$l(k_l) + H(P) \rightarrow  l({k}'_l) + H'(P_H) + X$
with lepton momenta $k_l$, ${k}'_l$ and space-like momentum transfer, $q={k}_l-{k}'_l$ with $Q^2=-q^2$.
The momenta of the incoming and outgoing hadrons are denoted as $P$, $P_H$.
The computation of SIDIS observables in perturbation theory can nowadays be performed with an unprecedented accuracy, thanks to the remarkable theoretical developments in the calculation of multi-loop and multi-leg scattering processes. 
In addition, resummation techniques applicable in particular kinematical limits allow, e.g. to obtain approximate predictions based on the summation of large threshold logarithms to all orders, see refs.~\cite{Cacciari:2001cw,Anderle:2012rq,Anderle:2013lka,Abele:2021nyo,Abele:2022wuy}.

The longitudinal momentum distribution of the final state hadron is
sensitive to the fragmentation functions and depends on scaling variables $x$, $y$ and $z$,
\begin{eqnarray}
\frac{d^2\sigma_{e^{-}H}}{dxdydz} =\frac{2\pi y \alpha_e^{2}}{Q^4}L^{\mu\nu}(k_l,k'_l,q)W_{\mu\nu}(q,P,P_H)
\, ,
\end{eqnarray}
where $\alpha_e$ is the fine structure constant, $x=\frac{Q^2}{2 P \cdot q}$ is the Bjorken variable, $y=\frac{P\cdot q}{P\cdot k_l}$ the fraction of the initial energy transferred to the hadron 
and $z= \frac{P \cdot P_H}{P \cdot q}$ the scaling variable corresponding to the fragmenting hadron.
The leptonic tensor reads 
$L^{\mu\nu}=2{k}_l^{\mu}{k}_{l}^{\prime\nu}+2{k}_{l}^{\prime\mu}{k}_l^{\nu} -Q^2 g^{\mu\nu}$, while the hadronic tensor $W_{\mu\nu}$ can be expressed in terms of SFs $F_I(x,z,Q^2), I=1,2$, as
\begin{eqnarray}
W_{\mu\nu} = F_1 (x,z,Q^2)T_{1,\mu \nu} + F_2 (x,z,Q^2) T_{2,\mu \nu}
\, ,
\end{eqnarray}
with tensors 
$T_{1,\mu\nu}$ = $-g_{\mu \nu}+\frac{q_\mu q_\nu}{q\cdot q}$ 
and 
$T_{2,\mu\nu}= \frac{1}{P.q}(P_\mu - \frac{P\cdot q}{q\cdot q }q_\mu)(P_\nu - \frac{P\cdot q}{q\cdot q} q_\nu)$.
QCD factorization allows us to express the SFs as
\begin{align}
\label{eq:SFdef}
F_I &= x^{I-1}\sum_{a,b}\int_x^1 \frac{dx_1}{x_1} f_a(x_1,\mu_F^2) \int_z^1 \frac{dz_1}{z_1} D_b(z_1,\mu_F^2) 
\nonumber \\
    & \times \mathcal{F}_{I,ab}\left( \frac{x}{x_1}, \frac{z}{z_1}, \mu_F^2, Q^2\right )
\, ,
\end{align}
where $f_a$ and $D_b$ denote the PDFs and the FFs, subject to summation over initial state partons `$a$' from the incoming hadron and final state partons `$b$' that fragments into the observed hadron. 
The coefficient functions (CFs) $\mathcal{F}_{I,ab}$ can be computed in a perturbative expansion in powers of the strong coupling $a_s(\mu_R^2) = \frac{\alpha_s(\mu_R^2)}{4 \pi}$,
\begin{eqnarray}
\label{eq:as-exp}
{\cal F}_{I,ab}&=& \sum_{i=0}^\infty\, a_s^i(\mu_R^2)\, {\cal F}_{I,ab}^{(i)}(\mu_R^2)
\, ,
\end{eqnarray}
with $\mu_F$, $\mu_R$ the factorisation and renormalisation scales.

To date, exact results for CFs are available only at the next-to-leading order (NLO) in QCD~\cite{Altarelli:1979kv,Furmanski:1981cw}, 
covering all parton combinations $ab = qq, \bar{q} \bar{q}, qg, \bar{q}g, gq, g\bar{q}$. 
In this letter, we present the first results at next-to-next-to leading order (NNLO) for the non-singlet channels $ab=q q, \bar{q} \bar{q}$, 
i.e., for SIDIS processes with an initial quark or an anti-quark, which fragments into the observed final state hadron.
The results are complete to leading color $N_c$ for QCD as an $SU(N_c)$ gauge theory.
Previously, refs.~\cite{Abele:2021nyo}, \cite{Abele:2022wuy} have derived the threshold enhanced logarithms using resummed results for these partonic channels ($ab=q q ,\bar{q} \bar{q}$)
at NNLO and even up to next-to-next-to-next-to leading order (N$^3$LO). 
We confirm that the resummation of ref.~\cite{Abele:2021nyo} correctly predicts the dominant threshold logarithms.
 
The computation of the CFs, $\mathcal{F}_{I,ab}$ in perturbative QCD in powers of $a_s$, see eq.~(\ref{eq:as-exp}), starts from the parton level cross sections denoted by $d\hat{\sigma}_{I,ab}$, appropriately set up with the corresponding projectors $\mathcal{P}^{\mu\nu}_I$,
\begin{eqnarray}
\label{eq:parton-crs}
d\hat{\sigma}_{I,ab} = \frac{\mathcal{P}_I^{\mu\nu}}{4\pi}\int \text{dPS}_{X+b}\, \overline{\Sigma}|{M}_{ab}|^{2}_{\mu\nu}\, \delta\Big(\frac{z}{z_1}-\frac{p_a \cdot p_b}{p_a\cdot q}\Big) 
\, ,
\end{eqnarray}
where $|M_{ab}|^2$ is the squared amplitude for the process $a(p_a)+\gamma^{*}(q)\rightarrow b(p_b) + X $, with the parton `$b$' tagged to fragment into hadron $H'$.  $dPS_{X+b}$ stands for the phase space of the final state consisting of $X$ and the fragmenting parton $b$ and $\overline{\Sigma}$ denotes the average over initial and summation over final state spin/polarization and color quantum numbers.
In $D$ dimensions, the projectors $\mathcal{P}^{\mu\nu}_{I}$ are given by 
$\mathcal{P}^{\mu\nu}_1$=$\frac{1}{(D-2)}(T^{\mu\nu}_1+2xT^{\mu\nu}_2)$ and $\mathcal{P}^{\mu\nu}_2$=$\frac{2x}{(D-2)x_1}(T^{\mu\nu}_1+2x(D-1)T^{\mu\nu}_2)$. 
The partonic scaling variables are $x_1=\frac{p_a}{P}$ for the momentum fraction carried by the initial parton `$a$' of incident hadron $H$ and $z_1=\frac{P_H}{p_b}$ for the corresponding fraction of hadron $H'$ with respect to the final state parton `$b$'.

Beyond leading order in perturbation theory, both ultraviolet (UV) and infrared divergences resulting from soft and collinear partons appear in the computation of $d\hat \sigma_{I,ab}$, and we work in $D=4+\varepsilon$ dimension to regulate them.
The ultraviolet divergences are removed by renormalisation of the strong coupling at the scale $\mu_R$, while all infrared divergences cancel among virtual and real emission processes, except for the collinear divergences related to the `$ab$' partons in the initial state and the fragmentation into hadrons. 
The structure of those divergences is described by mass factorisation in QCD. 
The partonic cross sections in eq.~(\ref{eq:parton-crs}) factorise into the Altarelli-Parisi (AP) kernels 
$\Gamma_{c\leftarrow a}$ of PDFs and $\tilde \Gamma_{b\leftarrow d}$ of FFs, which capture the collinear divergences in $1/\varepsilon$,
and the CFs (${\cal F}_{I,cd}$), which are finite as $\varepsilon \rightarrow 0$.
Mass factorisation at the scale $\mu_F$ reads
\begin{eqnarray}
\label{eq:massfact}
\frac{d\hat \sigma_{I,ab}(\varepsilon)}{x'^{I-1}} = 
\Gamma_{c\leftarrow a}(\mu_F^2,\varepsilon) \otimes 
{\cal F}_{I,cd}(\mu_F^2,\varepsilon) \tilde \otimes 
\tilde \Gamma_{b\leftarrow d}(\mu_F^2,\varepsilon)
\, ,
\, 
\end{eqnarray}  
where $x'=x/x_1$, summation over $c,d$ is implied and $\otimes$ and $\tilde \otimes$ denote
convolutions over scaling variables corresponding to the PDFs and FFs, respectively,
cf. eq.~(\ref{eq:SFdef}).
The AP kernels are determined by the evolution equations for PDFs and FFs in terms of space- and time-like splitting functions, fully known to third order in $a_s$~\cite{Moch:2004pa,Vogt:2004mw,Almasy:2011eq,Chen:2020uvt}.
Due to the collinear poles in $1/\varepsilon$ of the AP kernels, the extraction of the CFs (${\cal F}_{I,ab}$) to NNLO accuracy from the convolutions in eq.~(\ref{eq:massfact}) requires the computation of NLO $d\hat \sigma_{I,ab}$ up to positive powers of $\varepsilon$.

At NLO, the partonic cross sections in eq.~(\ref{eq:parton-crs}) get contributions from 
the one-loop corrections to the Born process $\gamma^*+q (\bar{q}) \rightarrow q (\bar{q})$
and the real emission $\gamma^* + q (\bar{q}) \rightarrow q (\bar{q}) + g$. 
In addition, there is the gluon-initiated channel $\gamma^* + g \rightarrow q + \bar{q}$.
In the new NNLO computation of $d\hat \sigma_{i,qq}$, we restrict ourselves to the quark flavor 
non-singlet case, i.e., we consider the following three classes, namely, 
two-loop corrections to $\gamma^* + q(\bar{q}) \rightarrow q(\bar{q})$,
one-loop contributions to the single gluon real emission $\gamma^* + q (\bar{q}) \rightarrow q (\bar{q}) + g$, 
and double real emissions $\gamma^* + q(\bar{q}) \rightarrow q(\bar{q}) + g + g$.  

One- and two-loop virtual corrections to the Born process can be obtained using the vector form factor 
which is known to fourth order in $a_s$~\cite{Lee:2022nhh}.  
For the real-virtual and double real emission processes we follow
the standard diagrammatic approach. We generate Feynman diagrams using \textsc{QGRAF} \cite{Nogueira:1991ex} 
and use a set of in-house routines in \textsc{FORM} \cite{Kuipers:2012rf,Ruijl:2017dtg}, which convert the output of
\textsc{QGRAF} into a suitable format that allows one to use the Feynman rules and to perform Lorentz contractions, color as well as Dirac algebra.  
The real-virtual processes contain one-loop and two-body phase-space integrals, while double real emission processes contain three-body phase-space integrals. 
The phase-space integrals are performed with the constraint 
$\frac{z}{z_{1}} = \frac{p_a \cdot p_b}{p_a\cdot q}$, because of which, the computation of these phase-space integrals is technically challenging 
compared to the fully inclusive ones. 
Using the method of reverse unitarity~\cite{Anastasiou:2003gr,Anastasiou:2012kq},
we convert all the phase-space integrals into loop integrals and apply integration-by-parts identities (IBP) 
\cite{Chetyrkin:1981qh,Laporta:2001dd} to reduce them to a smaller number of the master integrals (MI).   
In the reverse unitary method, we replace the on-shell Dirac delta functions by the 
corresponding propagators. 
The constraint ${z\over z_{1}}= \frac{p_a \cdot p_b}{p_a \cdot q}$ is imposed by an additional delta function $\delta \left(z' - \frac{p_a \cdot p_b}{p_a \cdot q} \right)$, 
where $z'=\frac{z}{z_1}$, which is then replaced by a propagator-like term 
$-\frac{1}{\pi}\text{Im}(1/(z' - \frac{p_a \cdot p_b}{p_a \cdot q} + i \epsilon))$ 
with $p_2 = p_1 + q - k_1$ or $p_2 = p_1 + q - k_1 - k_2$ 
for two- or three-body final states. 
We perform the IBP reduction with the \textsc{Mathematica} package \textsc{LiteRed} \cite{Lee:2013mka} and obtain at the NNLO level at leading $N_c$, 7 MIs for both the real-virtual and the double real emission sub-processes.
We note that the computation of the double real emission process $\gamma^* + q(\bar{q}) \rightarrow q(\bar{q}) + q + \bar{q}$ as a part of the flavor pure-singlet SFs (not considered in this paper)
requires an additional 13 MIs for the color non-planar contributions, i.e. those suppressed as $1/N_c$.

The calculation of the MIs poses one of the primary challenges in this computation. 
We have followed two different approaches, applying either parametric integration (PI)~\cite{Matsuura:1988sm,Zijlstra:1992qd,Rijken:1996ns,Ravindran:2003um}
or differential equations (DE)~\cite{Kotikov:1990kg,Argeri:2007up,Remiddi:1997ny,Henn:2013pwa,Ablinger:2015tua}.
In the PI method, a convenient choice of the Lorentz frame helps to solve the integrals.
For example, in the case of a two-body phase-space, we use the center-of-mass (COM) frame of the photon and the incoming parton, so that only the one-loop integrals need to be done, which reduce to hypergeometric functions of the scaling variables
$x'$, $z'$ and $\varepsilon$.
The three-body phase-space is conveniently integrated in the COM frame of two outgoing partons that do not fragment and we need to perform integrals over one of the parametric variables and two angles. Upon expansion in $\varepsilon$ the three-body phase-space integrals then reduce to multiple polylogarithms (MPLs) and Nielsen polylogarithms.

In the second approach, we have used the method of differential equations
to evaluate the MIs. 
The corresponding system of differential equations has been generated for each topology by taking derivatives with respect to the independent variables $x'$ and $z'$, using \textsc{LiteRed} for the differentiation and further IBP reduction.
We have obtained a system without any coupled differential equations, which can easily be arranged in an upper-triangular form, and allows for the bottom-up approach to solve the MIs one by one.
While solving for each MI, we also perform a Taylor series expansion in $\varepsilon$ and obtain the results
in terms of generalized harmonic polylogarithms (GPLs).
The boundary conditions for the MIs are either derived from regularity conditions or by calculating explicitly the threshold limit.
We have used \textsc{HarmonicSums} \cite{Ablinger:2010kw,Ablinger:2011te,Ablinger:2014rba}  and \textsc{PolyLogTools} 
\cite{Duhr:2019tlz} in various intermediate steps
of the computation including the conversion of MPLs and Nielsen polylogarithms into GPLs.
We have cross-checked the results obtained from both approaches and found perfect agreement.

The phase-space integrals in $D=4+\varepsilon$ generally result into functions of the form 
$$
(1-x')^{-1 + a \frac{\varepsilon}{2}} (1-z')^{-1 + b \frac{\varepsilon}{2}} f_1(x',z',\varepsilon)
\, ,
$$ 
which contain simple poles at the thresholds i.e. $x' \rightarrow 1$ and $z' \rightarrow 1$ for 
integers $a$, $b$.
In addition, we find  singularities at $x'=z'$ and at $x'+z'=1$ in the form
$$
\frac{(1-x')^{a \varepsilon}   (1-z')^{b \varepsilon}(z'-x')^{c \varepsilon}(1-z'-x')^{d \varepsilon} }{(1-x') (1-z')(z'-x')(1-z'-x')} f_2(x',z',\varepsilon)
\, .
$$
The $f_i(x',z',\varepsilon)$ are regular functions in the threshold limits $x'\rightarrow 1$ and/or $z'\rightarrow 1$. 
The sign of the imaginary part of the term $(z'-x')^{a \varepsilon}$ depends on whether $x'>z'$ or $x'<z'$,
while for the term $(1-z'-x')^{b \varepsilon}$, it depends on whether $x'+z'>1$ or $x'+z'<1$.   We introduce the identities $\theta(x'-z')+\theta(z'-x')=1$ and $\theta(1-x'-z')+
\theta(z'+x'-1)=1$ to separate the different sectors.
Note that these scaling variables should be understood as $x'-i\epsilon$ and $z'-i\epsilon$ with Feynman's $i\epsilon$ prescription.
The divergences resulting from the threshold region can be isolated by using the following identity
\begin{align}
(1-w)^{-1 + n \varepsilon} = \frac{1}{n \varepsilon} \delta(1-w) + \sum_{k=0}^\infty \frac{(n \varepsilon)^k}{k!}  \mathcal{D}_{w,k} \,,
% \left[ (1-w)^{-1 + n \varepsilon} \right]_+ \,.
\end{align}
where $w=x',z'$ and 
$\mathcal{D}_{w,k} = \left[ \frac{\log^k (1-w)}{(1-w)} \right]_+$
denote the usual `plus'-distributions, as defined by
\begin{equation}
 \int_0^1 dw ~ g_+(w) ~ f(w) = \int_0^1 dw ~ g(w) ~ (f(w)-f(1)) 
 \, .
\end{equation}
Due to the presence of these singularities, it is henceforth necessary to compute the phase-space 
integrals in closed form in $\varepsilon$, or at least partially. 
At NLO level, the leading double poles ($1/\varepsilon^2$) from the virtual contributions and the real emission diagrams  
cancel each other and the remaining collinear divergence ($1/\varepsilon$) is removed by mass factorisation with the AP kernels $\Gamma$ and $\tilde \Gamma$, see eq.~(\ref{eq:massfact}).  
At NNLO level, the leading $1/\varepsilon^4$ and $1/\varepsilon^3$ terms cancel among
the pure virtual, real-virtual and double real emission contributions.  
All remaining $1/\varepsilon^2$ and $1/\varepsilon$ terms cancel against those of the AP kernels during mass factorization.

After mass factorization given by eq.~(\ref{eq:massfact}) we obtain the finite CFs 
${\cal F}_{I}^{(i)}$ ($\equiv {\cal F}_{I,ab}^{(i)}$ as a short-hand now) 
for $i=0,1,2$.
The Born contribution ${\cal F}_I^{(0)}$ is proportional to $\delta(1-x')\delta(1-z')$ 
and our NLO results for ${\cal F}_I^{(1)}$ are in complete agreement
with refs.~\cite{Anderle:2012rq,Altarelli:1979kv}. 
It is convenient to separate the dependence on distributions $\delta(1-x'), \delta(1-z'), \mathcal{D}_{x',k}, \mathcal{D}_{z',k}$ and on regular functions in $x'$, $z'$ such as MPLs 
as follows,
\begin{equation}
    \label{eq:FI-def}
    {\cal F}_I^{(i)} ={\cal F}_{I,2}^{(i)} + {\cal  F}_{I,1}^{(i)} +{\cal F}_{I,r}^{(i)}
    \, .
\end{equation}
At NNLO, the first term ${\cal F}_{I,2}^{(2)}$ contains only distributions in $x'$ and $z'$,
often called soft-plus-virtual (SV) terms  and is in agreement with ref.~\cite{Abele:2021nyo}, which provides the first verification of those results from an explicit computation.  
The remaining contributions in eq.~(\ref{eq:FI-def}), consisting of single distributions denoted by ${\cal F}_{I,1}^{(2)}$ and pure regular terms ${\cal F}_{I,r}^{(2)}$, are new.  
As the latter are too lengthy, we present here only the single distributions ${\cal F}_{1,1}^{(2)}$ and introduce the following abbreviations for compact presentation: 
$
\xb = (1-x')\,, \zb = (1-z')\,, 
\xt = (1+x')\,, \zt = (1+z')\,, 
\delta_{\xb} = \delta (1-x') \,,
\delta_{\zb} = \delta (1-z') \,,
l_x=\log(x')\,,  l_z=\log(z')\,,  
l_{\bar{x}} = \log(1-x')\,,  
l_{\bar{z}} = \log(1-z')\, $. 
The full results for ${\cal F}_{I}^{(i)}$ in eq.~(\ref{eq:FI-def}) are attached to this publication as a supplementary file. 
We also note that, although our NNLO results consist of the leading color contributions, 
we retain their full Casimir structure in the presentation, since all 
sub-leading color contributions (to be presented in future work) are proportional 
to $C_F (C_A - 2 C_F)$.
Thus, the NNLO result for the single distributions of ${\cal F}_{1}^{(2)}$ reads:
\begin{widetext}
\begin{align}
{\cal F}_{1,1}^{(2)} &= C_F^2 \Big[
 \delta_{\bar{x}} \Big\{
        2 \logz^2 (1 - 4\zb)
        +4  (1 - 8\zb)
        -8 \lithzbar \zt
        +\frac{25}{3}  \logz^3 \zt
        -4 \logz \logzbar^2 \zt
        -4 \logzbar^3 \zt
        +52 \sotzbar \zt
        +\litwzbar (4  (1 - 6\zb)
\nn\\& ~~  
        +40 \logz \zt
        )
        +\iomz \big(
                8 \lithzbar
                -64 \litwzbar \logz
                -\frac{40}{3}  \logz^3
                +12 \logz \logzbar^2
                -88 \sotzbar
                +\logzbar \big(
                        -8 \litwzbar
                        -12 \logz^2
                \big)
                +\logz \big(
                        -64
                        +24 \zeta_2
                \big)
        \big)
\nn\\& ~~
        +\logzbar \big(
                14
                +24 \zt
                +4 \logz  (1 - 2\zb)
                +8 \litwzbar \zt
                +10 \logz^2 \zt
                +16 \zt \zeta_2
        \big)
        +\logz \big(
                -2
                +38 \zt
                -16 \zt \zeta_2
        \big)
        +8 \zb \zeta_2
        -16 \zt \zeta_3
\Big\}
% % % % % % % % % % % 
\nn\\& ~~
+ {\mathcal{D}_{x,0}}  \big\{
        12
        +24 \zt
        +4 \logz   (1 - 3\zb)
        +12 \litwzbar \zt
        +16 \logz^2 \zt
        -4 \logz \logzbar \zt
        -12 \logzbar^2 \zt
        -\iomz \big(
                 16 \litwzbar
                +24 \logz^2
                -16 \logz \logzbar
        \big)
        +16 \zt \zeta_2
\big\}
% % % % % % % % % % % 
\nn\\& ~~
+ {\mathcal{D}_{x,1}}  \big\{ (4 \logz \zt
-24 \logzbar \zt ) 
\big\}
% 
% % % % % 
-12 \zt ~ {\mathcal{D}_{x,2}}
+\delta_{\bar{z}} \Big\{
        -4
        -48 \xb
        -2 \logx^2
        +\frac{11}{3}  \logx^3 \xt
        +16 \logx \logxbar^2 \xt
        -4 \logxbar^3 \xt
        -24 \sotxbar \xt
\nn\\& ~~        
        +\litwxbar (4 + 8 \xb
        -12 \logx \xt
        )
        +\iomx \big(
                -8 \lithxbar
                +16 \litwxbar \logx
                -4 \logx^3
                -28 \logx \logxbar^2
                +48 \sotxbar
                +\logxbar \big(
                        8 \litwxbar
                        +32 \logx^2
                \big)
\nn\\& ~~                
                +\logx \big(
                        64
                        +32 \zeta_2
                \big)
        \big)
        +\logxbar \big(
                14
                +26 \xt
                +4 \logx
                -20 \logx^2 \xt
                +16 \xt \zeta_2
        \big)
        +\logx \big(
                -8
                -34 \xt
                -20 \xt \zeta_2
        \big)
        +8 \xb \zeta_2
        -16 \xt \zeta_3
\Big\}
% % % % % % % % % % % 
\nn\\& ~~ 
+  {\mathcal{D}_{z,0}} \big\{
        12
        +28 \xt
        +4 \logx (1+\xb)
        -4 \litwxbar \xt
        -12 \logx^2 \xt
        +28 \logx \logxbar \xt
        -12 \logxbar^2 \xt
        +\iomx \big(
                16 \litwxbar
                +16 \logx^2
                -48 \logx \logxbar
        \big)
        +16 \xt \zeta_2
\big\} 
% % % 
\nn\\& ~~ 
% % % % 
+ {\mathcal{D}_{z,1}} \Big\{  
-\frac{32}{\bar{x}} \logx
+20 \logx \xt
-24 \logxbar \xt
\Big\}
% % % % % % 
-12 \xt {\mathcal{D}_{z,2}} 
\Big]
% 
% 
% % % % % % % % % % % % % % % % % % % % % % % % % % % 
% 
+ C_A C_F   \Big[
 \delta_{\bar{x}} \Big\{
         4 \litwzbar (1 - 2 \zt)
        +\frac{2}{3} \logz^2   (3 - 11 \zt)
        +\frac{1}{27} (396 + 179 \zt)
\nn\\&~~        
        +\frac{1}{9} \logz (1+70 \zb)
        -4 \lithzbar \zt
        -\frac{5}{3}  \logz^3 \zt
        +\frac{11}{3}  \logzbar^2 \zt
        +6 \sotzbar \zt
        +\iomz \big(
                6 \litwzbar
                +8 \lithzbar
                +\frac{62}{9}  \logz
                +\frac{49}{6}  \logz^2
                +\frac{10}{3}  \logz^3
                +6 \logz \logzbar
\nn\\&~~                
                -12 \sotzbar
        \big)
        +\logzbar \Big(
                 \frac{4}{9} (15 - 41 \zt)
                -6 \logz \zt
                +4 \zt \zeta_2
        \Big)
        -\frac{4}{3} (3 + 4 \zt) \zeta_2
        -14 \zt \zeta_3
\Big\}
% % 
+ {\mathcal{D}_{x,0}}  \Big\{
         \frac{2}{9}  (39 - 82 \zt)
        -4 \litwzbar \zt
        -6 \logz \zt
\nn\\&~~        
        -2 \logz^2 \zt
        +\frac{22}{3}  \logzbar \zt
        +\iomz \big(
                8 \litwzbar
                +6 \logz
                +4 \logz^2
        \big)
        +4 \zt \zeta_2
\Big\} 
+\frac{22}{3} \zt  {\mathcal{D}_{x,1}} 
% 
% % % % % % % % % % % % % % % 
% 
+  \delta_{\bar{z}} \Big\{
        \frac{46}{3}
        +\frac{197}{27}  \xt
        +8 \lithxbar \xt
        +\frac{55}{6}  \logx^2 \xt
        +\frac{11}{3}  \logxbar^2 \xt
\nn\\&~~        
        +2 \sotxbar \xt
        +\litwxbar \Big(
                -4 \logx \xt
                -\frac{4}{3} (3+4 \xt)
        \Big)
        +\logx \Big(
                \frac{1}{3} (-13+77 \xt)
                -4 \xt \zeta_2
        \Big)
        +\iomx \Big(
                -16 \lithxbar
                -\frac{83}{6} \logx^2
\nn\\&~~
                +\frac{2}{3} \litwxbar (13+12 \logx)
                +\frac{70}{3}  \logx \logxbar
                -4 \sotxbar
                +\logx \Big(
                        -\frac{116}{3}
                        +8 \zeta_2
                \Big)
        \Big)
        +\logxbar \Big(
                -\frac{44}{3}  \logx \xt
                -\frac{4}{9} (-15+41 \xt)
                +4 \xt \zeta_2
        \Big)
\nn\\&~~        
        -\frac{4}{3} (3+4 \xt) \zeta_2
        -14 \xt \zeta_3
\Big\}
% % % % 
+  {\mathcal{D}_{z,0}}  \Big\{
        4 \litwxbar \xt
        -\frac{44}{3}  \logx \xt
        +\frac{22}{3}  \logxbar \xt
        +\frac{26}{9} (3-7 \xt)
        +\iomx \big(
                -8 \litwxbar
                +\frac{70}{3} \logx
        \big)
        +4 \xt \zeta_2
\Big\} 
\nn\\&~~        
+\frac{22}{3} \xt  {\mathcal{D}_{z,1}} 
\Big]
% 
% % % % % % % % % % % 
+ \frac{1}{3} C_F n_F \Big[
\delta_{\bar{x}} \Big\{
        4
        -\frac{2}{3} \iomz \logz (10+3 \logz)
        -\frac{74}{9}  \zt
        +\logz^2 \zt
        -2 \logzbar^2 \zt
        +\frac{8}{3} \logzbar (-3+4 \zt)
        +\frac{2}{3} \logz (-12+11 \zt)
        +4 \zt \zeta_2
\Big\}
\nn\\&~~
+  {\mathcal{D}_{x,0}} \Big\{
         \frac{32}{3}  \zt
        -8
        -4 \logzbar \zt
\Big\} 
-4 \zt {\mathcal{D}_{x,1}}
+\delta_{\zb} \Big\{
         4 \litwxbar \xt
        -4
        -\frac{38}{9} \xt
        -5 \logx^2 \xt
        -2 \logxbar^2 \xt
        +2 \logx (2-7 \xt)
        +\iomx \big(
                 20 \logx
                +10 \logx^2
\nn\\&~~                 
                -8 \litwxbar
                -16 \logx \logxbar
        \big)
        +\logxbar \Big(
                 \frac{32}{3} \xt
                -8
                +8 \logx \xt
        \Big)
        +4 \xt \zeta_2
\Big\}
+  {\mathcal{D}_{z,0}} \Big\{
         \frac{32}{3} \xt
        -8
        -\frac{16}{\xb}   \logx
        +8 \logx \xt
        -4 \logxbar \xt
\Big\} 
-4 \xt   {\mathcal{D}_{z,1}} 
\Big]  \,.
\end{align}
\end{widetext}
% 
%----------------------------------------------------------------------------------------------------%
% 
where $S_{n,p}(z)$= ${(-1)^{n+p-1} \over (n-1)! p!} \int_0^1 {dy \over y} \log^{n-1}(y) \log^p(1-z y) $, $Li_n(z)$ = $S_{n-1,1}(z)$, 
and $\zeta_n$ denote values of the Riemann zeta-function.

A few points are in order. As a check of our results, we have reproduced the inclusive SF results of refs.~\cite{Zijlstra:1992qd,vanNeerven:1991nn} for the channels considered by integrating over the scaling variable $z'$, including all the
scale-dependant terms.
The terms that are proportional to the Dirac delta functions and `plus' distributions are in complete agreement with ref.~\cite{Abele:2021nyo}, which uses the framework of threshold resummation. 
Our results from an explicit Feynman diagrammatic approach confirm those predictions including the next-to-SV (NSV) terms quoted in ref.~\cite{Abele:2021nyo}.

\begin{figure}[!ht]
\includegraphics[width=0.538\textwidth,height=0.548\textwidth]
{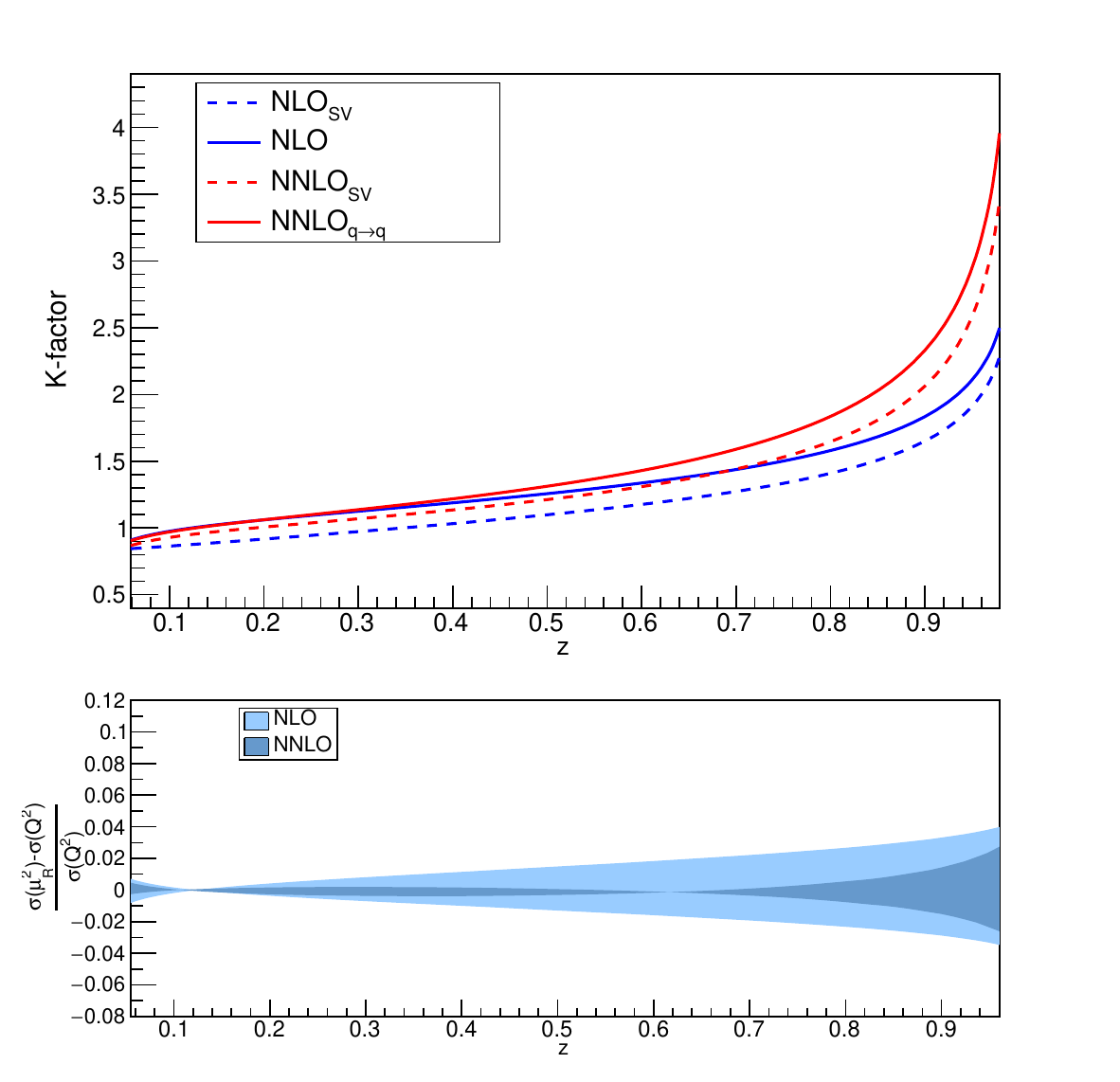}
\caption{The upper panel contains the $K$-factor as a function of $z$ for the NLO and NNLO results, using kinematics of the EIC at $\sqrt{s}=140$ GeV and different approximations:
the SV terms at NLO (blue dashed), full NLO (blue solid), 
SV terms at NNLO (red dashed) and full (non-singlet, leading color) NNLO (red solid). 
The lower panel contains the corresponding uncertainty due to the renormalisation scale variation in the range $\mu_R^2 \in [Q^2/2,2Q^2]$.}\label{fig:1}
\end{figure}

In the following we illustrate the numerical impact of the new results for the  ${\cal F}_{1,qq}$ CF for a future EIC with a COM energy $\sqrt{s}$ = 140~GeV.
We have convoluted the CFs with a set of order independent, but sufficiently realistic model distributions for both, PDFs and FFs, 
\begin{eqnarray}
xq(x,\mu_F^2)&=& 0.6 x^{-0.3}(1-x)^{3.5}(1 + 5.0x^{0.8})\, ,\nonumber \\
xg(x,\mu_F^2)&=& 1.6 x^{-0.3}(1-x)^{4.5}(1 - 0.6x^{0.3})
\, .
\end{eqnarray}
In Fig.~\ref{fig:1} we present the $K$-factor, defined by the ratio
$K$=N$^i$LO/LO for $i=1,2$ as function of $z$ 
for EIC after integrating $x$ between $0.1$ to $0.8$ and $y$ between $0.1$ to $0.9$. 
We also show the variation of the renormalisation scale $\mu_R^2$ 
in the range $\mu_R^2 \in [Q^2/2,2Q^2]$, keeping $\mu_F^2$ = $Q^2$ fixed. 
We use $n_F=5$ for the number of active quarks, the value of $\alpha_e$=1/128 and for the strong coupling constant $\alpha_s(M_Z)= 0.120$ at NLO and $\alpha_s(M_Z)= 1.118$ at NNLO.

We find, for example at $z=0.5$, that the new NNLO contributions increase the $K$-factor from $1.26$ at NLO to $1.32$ at NNLO level, while reducing the renormalisation 
scale dependence from variations of $\mu_R^2$ by a factor of 2 around $Q^2$  
from $\{1.51\%, -1.31\%\}$ at NLO to $\{-0.32\%, 0.06\%\}$ at NNLO.

In this letter, we report the first results for the complete NNLO QCD corrections to the quark-initiated SIDIS process, 
based on calculating the complete set of Feynman diagrams including all relevant interference terms.
However, as mentioned earlier, we have selectively excluded the flavor pure-singlet parts of the double real emission process $\gamma^* + q(\bar{q}) \rightarrow q(\bar{q}) + q + \bar{q}$.
The contributions from these diagrams along with the detailed 
description of the computational method will be presented in a future publication. 

The primary challenge resides in the computation of the Feynman integrals, especially 
the phase-space integrals. 
To obtain them analytically, we have used state-of-the-art loop computation techniques,
namely the reverse unitarity method, the IBP reduction technique, and the method of differential equations. 
The parametric integration delivers the MIs in terms of MPLs and Nielsen polylogarithms, 
although their arguments are not simple, due to the presence of two variables, $x'$ and $z'$.
Simplification arise from a conversion to GPLs.
Our results have been exposed to a number of cross-checks, in particular the comparison of the SV and NSV limits with ref.~\cite{Abele:2021nyo}, finding complete agreement.

The new NNLO results display a moderate increase of the $K$-factor over a large range of kinematics, except for the threshold limits, where additional resummations need to be performed. At the same time, the residual scale uncertainties are significantly reduced.
As such, they mark a milestone in the precision study of the SIDIS process, and soon will play a very important role in shedding light on the physics of the hadron structure and the mechanism of fragmentation, facilitating new determinations of PDFs and FFs.

A \textsc{Mathematica} notebook with all results for the CFs ${\cal F}_I^{(i)}$ has been deposited at the preprint server {\tt https://arXiv.org} with the sources of this letter.
They are also available from the authors upon request.

%----------------------------------------------------------------------------------------------------%

\begin{acknowledgments}
\emph{Acknowledgements:}  
We thank W.~Vogelsang for discussions on ref.~\cite{Abele:2021nyo}.
This work has been supported through a joint Indo-German research grant by
the Department of Science and Technology (DST/INT/DFG/P-03/2021/dtd.12.11.21)
and Deutsche Forschungsgemeinschaft (project number 443850114).
S.M. acknowledges the ERC Advanced Grant 101095857 {\it Conformal-EIC}.

\emph{Note added:}
After completion of our work, ref.~\cite{Bonino:2024qbh} has appeared with the full NNLO QCD corrections to the SIDIS coefficient functions. 
Our results agree with those of ref.~\cite{Bonino:2024qbh} in all regions of $x'$and $z'$.
\end{acknowledgments}

\bibliographystyle{apsrev}
\bibliography{main}

\end{document}